\documentclass[pra,twocolumn,showpacs,amsmath,amssymb,superscriptaddress]{revtex4}

\usepackage{graphicx}
\usepackage{dcolumn}
\usepackage{bm}

\begin{document}

\title{Finite-Temperature Properties of Attractive 
Three-Component Fermionic Atoms in Optical Lattices}

\author{Kensuke Inaba}
\affiliation{
NTT Basic Research Laboratories, NTT Corporation, Atsugi 243-0198, Japan
}
\affiliation{
CREST, JST, Chiyoda-ku, Tokyo 102-0075, Japan
}

\author{Sei-ichiro Suga}%
\affiliation{
Department of Materials Science and Chemistry, University of Hyogo, Himeji 671-2280, Japan.
}
\affiliation{%
Department of Applied Physics, Osaka University, Suita, Osaka 565-0871, Japan
}%

\date{\today}

\begin{abstract}
We investigate the finite-temperature properties of attractive three-component (colors) fermionic atoms in
optical lattices using a self-energy functional approach. As the strength of the attractive interaction increases in
the low-temperature region, we observe a second-order transition from a Fermi liquid to a color superfluid
(CSF), where atoms from two of the three colors form Cooper pairs. In the strong attractive region, we observe
a first-order transition from a CSF to a trionic state, where three atoms with different colors form singlet bound
states. A crossover between a Fermi liquid and a trionic state is observed in the high-temperature region. We
present a phase diagram covering zero to finite temperatures. We demonstrate that the CSF transition temperature
is enhanced by the anisotropy of the attractive interaction.
\end{abstract}


\preprint{APS/123-QED}

\maketitle

The study of cold fermionic atoms has attracted considerable attention. Fascinating aspects of many-body effects have been revealed in various phenomena. 
A crossover between a BCS-type superfluid and Bose-Einstein condensation (BEC) was observed by controlling the strength of the attractive interaction using Feshbach resonances \cite{Regal,Bartenstein,Zwierlein,Kinast,Bourdel,Chin}. 
A superfluid-insulator transition was observed for $^6{\rm Li}$ fermionic atoms in an optical lattice \cite{Chin2006}. 
These phenomena with a highly tuned attractive interaction may be difficult to observe in condensed matter physics. 
One can expect further that the cold fermionic atoms show the novel phenomena beyond those observed in the condensed matter.
It has been shown that three-component (colors) fermionic atoms exhibit characteristic features. 
Their properties in optical lattices at zero temperature have been studied theoretically \cite{Honerkamp2004a,Honerkamp2004b,Rapp2007,Rapp2008}.
It was argued that for atoms with a weak attractive interaction two of the three colors form Cooper pairs, yielding a color superfluid (CSF). As the strength of the attractive interaction increases, there is a quantum phase transition from the CSF state to the trionic state, where three atoms with different colors form singlet bound states \cite{Rapp2007,Rapp2008}.

In contrast to the detailed investigations that have been undertaken at zero temperature, little information is available about the finite-temperature properties. 
Recently, fermionic atoms with a balanced population of three different hyperfine states were successfully created \cite{Ottenstein}. The temperature at which these atoms were realized was $T/T_{\rm F}\sim0.37$\cite{Ottenstein}, where $T_{\rm F}$ is the Fermi temperature. 
Studies of the finite-temperature properties are thus indispensable. 
In particular, the stabilities of the CSF and trionic phases against thermal fluctuations are important in terms of observing these novel states in experiments.

In this paper, we investigate fermionic atoms with three different colors $(\alpha=1,2,3)$ in optical lattices at zero and finite temperatures. Using a self-energy functional approach (SFA) \cite{Potthoff03a,Potthoff03b}, we elucidate characteristics of the CSF, trionic, and Fermi liquid states, and study the phase transition and crossover between them. 
In accordance with the conventional model for cold atoms in optical lattices \cite{Jaksch}, we set the nearest-neighbor hopping and the on-site attractive interaction between the atoms with different colors. 
Low-energy properties can be thus described by the attractive three-component Hubbard model, 
\begin{eqnarray}
{\cal H}&=-&\sum_{<i,j>}\sum_{\alpha=1}^{3}
       \left( t+\mu_\alpha \delta_{i,j} \right)c^\dag_{i\alpha} c_{j\alpha} 
   \nonumber \\
   &+& \frac{1}{2}\sum_{i}\sum_{\alpha\not=\beta} 
       U_{\alpha\beta} n_{i \alpha} n_{i \beta}
\label{eq_model},
\end{eqnarray}
where $c^\dag_{i\alpha} (c_{i\alpha})$ is the fermionic creation (annihilation) operator for the state with color $\alpha$ in the $i$th site and $n_{i\alpha}=c^\dag_{i\alpha}c_{i\alpha}$. 
$t$ denotes the nearest-neighbor hopping integral, 
$\mu_\alpha$ is the chemical potential for the atom with color $\alpha$, and 
$U_{\alpha\beta} (<0)$ is the attractive interaction between two atoms with colors $\alpha$ and $\beta$. 
The condition $\mu_\alpha= (1/2)(U_{\alpha\beta}+U_{\alpha\gamma})$ is imposed to maintain a half-filled system with a balanced population of three-component atoms. 
We assume a homogeneous optical lattice and neglect the confinement potential for a first approximation.

The SFA allows us to deal efficiently with zero- and finite-temperature properties concerning the phase transition driven by correlation effects. 
For Mott transitions in correlated electron systems, precise results have been obtained as regard to thermodynamic quantities, excitation spectra, and phase diagrams including the order of the transition \cite{Potthoff03a,Potthoff03b,Inaba05}. 
The SFA is based on the Luttinger-Ward variational method. 
This method enables us to introduce a proper reference system 
${\cal H}_{\rm ref}$, which has to include the same interaction term 
as that of the original Hamiltonian (\ref{eq_model}). 
A simple reference system is explicitly given by the following 
Hamiltonian, 
${\cal H}_{\rm ref}=\sum_i{\cal H}_{\rm ref}^{(i)}$, 
\begin{eqnarray}
{\cal H}_{\rm ref}^{(i)}=\sum_{\alpha=1}^3 \left( 
\epsilon_{c\alpha}c^\dag_{i\alpha} c_{i\alpha}+          
\epsilon_{a\alpha}a^{\dag}_{\alpha}a_{\alpha}\right)
          +\sum_{\alpha=1}^3\left(V_{\alpha}
          c^\dag_{i\alpha}a_{\alpha}+H.c.\right)\nonumber\\
+\frac{1}{2}\sum_{\alpha\not=\beta=1}^3\left(\Delta_{\alpha\beta}
          c^\dag_{i\alpha}c^\dag_{i\beta}+H.c.\right)
+\frac{1}{2}\sum_{\alpha\not=\beta=1}^3 
U_{\alpha\beta} n_{i \alpha} n_{i \beta},
\label{eq_ref_model}
\end{eqnarray}
where $a^{\dag}_{\alpha}(a_{\alpha})$ is the fermionic creation (annihilation) operator with color $\alpha$ connecting to the $i$th site in the original lattice. We then obtain the grand potential as 
\begin{eqnarray}
\Omega &=&\Omega_{\rm ref} +{\rm Tr}\ln
    \left[-(\omega+\mu-{\bf t}-{\boldsymbol \Sigma}_{\rm ref})^{-1}\right]
    \nonumber\\
    &-&{\rm Tr}\ln
    \left[-(\omega+\mu-{\bf t}_{\rm ref}-{\boldsymbol \Sigma}_{\rm ref})^{-1}\right],\label{eq:omega_SFA}
\end{eqnarray}
where $\Omega_{\rm ref}$ and ${\boldsymbol \Sigma}_{\rm ref}$ are the grand potential and the self-energy for the reference system, and 
${\bf t}_{\rm ref}$ and ${\bf t}$ are parameter matrices of the noninteracting terms of the reference Hamiltonial and the original Hamiltonian, respectively. 
By choosing the parameter matrix ${\bf t}_{\rm ref}$ to satisfy the condition 
$\partial\Omega/\partial {\bf t}_{\rm ref}=0$, we obtain a proper reference self-energy ${\boldsymbol \Sigma}_{\rm ref}$, which approximately describes the original correlated system. 
Under the conditions for $\mu_\alpha$ mentioned above, we can fix the parameters $\epsilon_{c\alpha}=0$ and $\epsilon_{a\alpha}=\mu_\alpha$. 
We choose a particular gauge for the pairing field so that 
only $\Delta\equiv\Delta_{12}=-\Delta_{21}$ components of the variational parameter $\Delta_{\alpha\beta}$ are nonzero \cite{Honerkamp2004a,Rapp2007}. 
From the condition $\partial\Omega/\partial\Delta=0$, we obtain $\Delta$ and 
can discuss whether the color superfluidity appears in the original system. 
To take the trionic state into account, we set $V_1=V_2=V_3\equiv V$. 
If the condition $\partial\Omega/\partial V=0$ is satisfied 
at $V=0$, the system is driven to the trionic phase expected to appear in a large $|U|/t$. 
By searching the optimized parameters $V$ and $\Delta$, we discuss 
a CSF-trion phase transition in the attractive three-component Hubbard model. 
In the following, we adopt the density of states for an infinite dimensional 
Bethe lattice, which is independent of color. 
The hopping integral is scaled as $t/\sqrt{d}$ to reach a meaningful limit \cite{Georges1996}. 
In this non-trivial limit, local correlation effects are known to cause  
various interesting phenomena in correlated electron systems, 
such as a Mott transition. 
Therefore, the infinite dimensional systems pave the way for a detailed study of correlation effects.

We calculate the CSF order parameter 
$\Phi=\langle c^\dag_{i1}c^\dag_{i2}\rangle$, 
the quasiparticle weight $Z_\alpha$, 
and the entropy per site $S/L=-\partial (\Omega/L)/\partial T$, 
where $L$ is the number of lattice sites. 
The quasiparticle weight $Z_\alpha$ is inversely proportional to the effective 
mass of the fermionic atom. 
We also calculate the single-particle excitation spectra (SPES) $\rho_{\alpha}(\omega/t)$. 
We do not consider the possibility of the trionic density-wave state induced by spatial fluctuations.

\begin{figure}[ht]
\includegraphics[width=7.5cm]{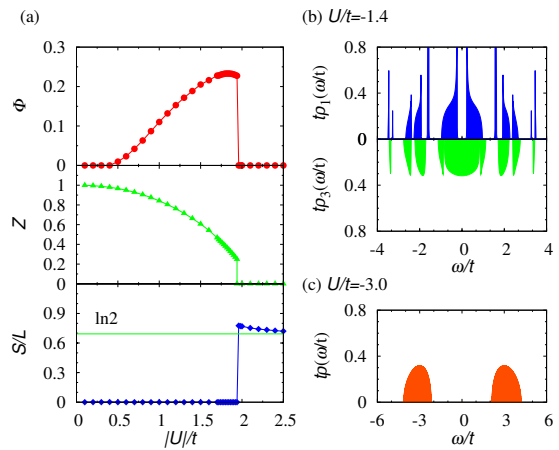}
\caption{(Color online) (a) $|U|/t$ dependence of the CSF order parameter $\Phi$, 
the quasiparticle weight $Z$, and the entropy per site $S/L$. 
(b) The single-particle excitation spectra of the $\alpha=1$ atoms $\rho_1(\omega/t)$ $[=\rho_2(\omega/t)]$ and those of the $\alpha=3$ atoms 
$\rho_3(\omega/t)$ at $U/t=-1.4$. 
(c) The single-particle excitation spectra $\rho(\omega/t)$ $[\equiv\rho_1(\omega/t)=\rho_2(\omega/t)=\rho_3(\omega/t)]$ at $U/t=-3.0$.
}\label{fig_ovs_zeroT}
\end{figure}

We begin our discussions with an isotropic interaction system at zero temperature. Figure \ref{fig_ovs_zeroT}\,(a) shows the results for $U_{12}=U_{23}=U_{31}\equiv U$. 
In the isotropic interaction system, $Z_\alpha$ becomes independent of $\alpha$. As $|U|/t$ is increased, the CSF order parameter $\Phi$ increases and then takes its maximum value at $U/t\sim-1.9$. 
At $U/t=-1.94$, $\Phi$ vanishes discontinuously, suggesting the first order 
transition. 
The quasiparticle weight $Z$ decreases significantly because of the renormalization effect and then drops suddenly to zero at $U/t=-1.94$.
The entropy $S/L$ jumps to $\sim \ln2$ at the same $|U|/t$. 
The residual entropy $S/L \sim \ln2$ in the large $|U|/t$ region indicates the formation of the localized trionic state at a site. 
The implications of this phenomenon are discussed later in this work. 
We thus conclude that the CSF-trion quantum phase transition occurs at $U/t=-1.94(\equiv U_{\rm c}/t)$, which is of the first order.

We investigate the CSF-trion transition with respect to the SPES. 
In Fig.\,\ref{fig_ovs_zeroT}(b) and (c), we show the SPES for 
$U/t=-1.4$ and $-3.0$. 
At $U/t=-1.4$, the SPES of colors 1 and 2 have a spectral gap around the Fermi energy $\omega=0$, which is caused by the superfluid order. 
Note that $\rho_1(\omega/t)=\rho_2(\omega/t)$. 
The SPES of color 3 has a Fermi liquid peak at $\omega=0$. 
The incoherent peaks at energies distant from the Fermi energy 
are seen not only in the Fermi liquid state but also in the CSF state. 
The results indicate that the attractive interaction induces a renormalized CSF state that is consistent with the decrease in the quasiparticle weight $Z$. 
In our calculations, leading excitation processes are taken account of, resulting in the multiple incoherent peaks. In the real system, continuous broad incoherent spectra are expected to appear instead of the multiple incoherent peaks.
On the other hand, the SPES of atoms with different colors are the same at $U/t=-3.0$, because the trionic state restores SU(3) symmetry broken in the CSF state \cite{Rapp2007}. 
The width of the spectral gap $\Delta_{\rm tri}$ is close to $2|U|/t$, which 
corresponds to the energy required to remove one atom from the trionic bound state.

In Refs. \cite{Rapp2007,Rapp2008}, the CSF-trion transition at $T=0$ was studied below half filling, using a Gutzwiller variational calculation combined with dynamical mean field theory. It was argued that the CSF-trion quantum phase transition is of the second order in contrast with our results. 
The SFA takes precise account of both the high- and low-energy properties of the self-energy. 
Actually, in the present system, we obtain the quasiparticle peaks in the CSF, the coherent peak in the Fermi liquid, and the incoherent peaks as shown in Fig.\,\ref{fig_ovs_zeroT}(b). 
However, in Gutzwiller variational calculations the correlation effects in the high-energy incoherent peaks are in principle neglected, although those in the low-energy coherent peak are included. 
When we evaluate the ground state energy, both the low- and high-energy contributions play important roles for obtaining precise results. 
It is considered that the accurate handling of the correlation effects in the SFA yields the first order quantum phase transition.

\begin{figure}[t]
\includegraphics[width=7.8cm]{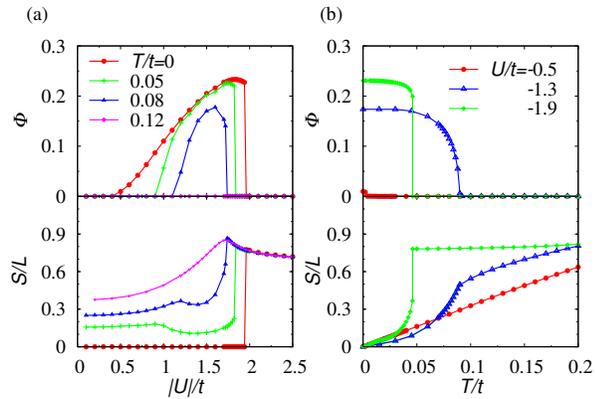}
\caption{(Color online) (a) $|U|/t$ dependence of the CSF order parameter $\Phi$ and the entropy per site $S/L$. 
(b) $T/t$ dependence of $\Phi$ and $S/L$. 
}\label{fig_ovs_finT}
\end{figure}
We next investigate the CSF order parameter and the entropy at finite temperatures. 
In Fig.\,\ref{fig_ovs_finT}(a), $\Phi$ and $S/L$ are shown as functions of 
$|U|/t$ for temperatures $T/t=0, 0.05, 0.08$, and $0.12$. 
As the interaction strength increases for $T/t=0.05$ ($0.08$), 
$\Phi$ increases continuously and $S/L$ shows a cusp at $|U|/t=0.96$ $(1.23)$. 
The results indicate a second order transition from the Fermi liquid to the CSF. 
As $|U|/t$ increases further, $\Phi$ and $S/L$ jump to zero and a finite value at $|U|/t=1.83$ $(1.73)$, respectively, which are evidences of the first order transition from the CSF to the trionic state. 
At $T/t=0.12$, $\Phi$ remains zero in overall $|U|/t$. 
For $S/L$, no discontinuity is observed but the peak around $U/t=-1.74$. 
The results demonstrate that there is a crossover from the Fermi liquid to the trionic state around $T/t=0.10$.

We investigate the temperature dependence of $\Phi$ and $S/L$ with the interaction strength fixed at $U/t=-0.5, -1.3$, and $-1.9$. 
The results are shown in Fig.\,\ref{fig_ovs_finT}(b).
For $U/t=-0.5$, the superfluidity vanishes except at $T/t\sim0$ and the entropy exhibits typical Fermi liquid behavior $S/L \propto T/t$. 
For $U/t=-1.3$, $\Phi$ decreases continuously with increasing $T/t$ and finally vanishes at $T/t=0.09$. 
We see that $S/L$ at $U/t=-1.3$ clearly changes its slope near the transition point $T/t\sim0.09$. 
At $T/t\lesssim0.09$, $S/L$ accords with the sum of the temperature dependence of the Fermi liquid and the superfluid, while $S/L$ at $T/t\gtrsim0.09$ exhibits a Fermi liquid behavior. 
For $U/t=-1.9$, $\Phi$ and $S/L$ show jumps at $T/t=0.04$, indicating a first order transition from the CSF to the trionic state. 
At $T/t\gtrsim 0.04$, $S/L$ remains almost constant, yielding zero specific heat. This feature is a manifestation of the trion gap $\Delta_{\rm tri}$.

In the trionic phase, the residual entropy at $T=0$ is $S/L \sim \ln2$ as seen in Fig.\,\ref{fig_ovs_zeroT}(a). 
The residual entropy is caused by pathological behavior typical in infinite dimensional systems. 
The effective hopping integral of the trion between neighboring sites can be derived as $t_{\rm eff}\propto t^3/U^2$ for a large $|U|/t$. 
In an infinite-dimensional system, $t_{\rm eff}$ becomes irrelevant \cite{Georges1996} and localized trions emerge. 
Accordingly, there are two degrees of freedom whether or not the trion is found in the site even at $T=0$, resulting in the residual entropy $\ln2$. 
In finite-dimensional systems, however, $t_{\rm eff}$ is relevant and forms a trionic Fermi liquid when $T \lesssim t_{\rm eff}$. 
For higher temperatures of $t_{\rm eff} < T < \Delta_{\rm tri}$, the fully mixed states of the thermally excited trions are expected to appear and the entropy $S/L$ takes $\ln2$ even in finite-dimensional systems. 
Our results for the trionic state may adequately describe the characteristic properties in this temperature region.

\begin{figure}[t]
\includegraphics[width=5.5cm]{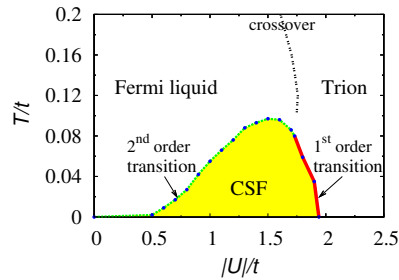}
\caption{(Color online) Phase diagram of the isotropic attractive three-component Hubbard model at half filling. 
}\label{fig_phase_twoD}
\end{figure}
By performing the same calculations, we obtain the phase diagram as shown in Fig.\,\ref{fig_phase_twoD}. 
We find two types of transition and a crossover, namely, the second order Fermi liquid-CSF transition, the first order CSF-trion transition, and the Fermi liquid-trion crossover. 
The crossover line is determined by the peak of $S/L$. 
The Fermi liquid-trion crossover line starts around the maximum CSF transition temperature $T_{\rm CSF}^{\rm max}$. This result is different from that schematically predicted in Ref. \cite{Rapp2008}, where the trion transition starts at the critical $|U_{\rm c}|/t$ of the CSF phase along the $T=0$ line. 
It is expected that these phase transitions and crossover can be detected by the temperature dependence of the entropy shown in Fig.\,\ref{fig_ovs_finT}(b). 
The trionic state may be detected by the change in the Fermi surface, since the coherent peak in the trionic state disappears as shown in Fig.\,\ref{fig_ovs_zeroT}(c)

Despite the high tunability of the optical lattice systems, it is difficult to cool the fermionic system because of the Pauli principle. 
%
%
To observe the CSF state, the CSF transition temperature has to be increased by controllable methods. To this end, we direct our attention to the effects of the anisotropy of the attractive interaction. In the experiment, the scattering lengths that control the interactions between the three atoms are different \cite{Ottenstein}. In addition, the respective scattering lengths are more tunable than temperature in experiments.

\begin{figure}[t]
\includegraphics[width=5.5cm]{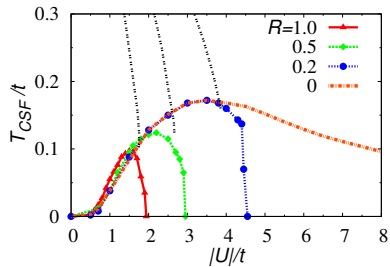}
\caption{
(Color online) CSF transition temperature $T_{\rm CSF}$ as a function of $|U|/t$ with $R=1.0, 0.5$,$0.2$, and $0$. The broken lines are the crossover lines for $R=1.0, 0.5$, and $0.2$ from left to right. 
}
\label{fig_phase_R}
\end{figure}
We set $U\equiv U_{12}$ and $U'\equiv U_{13}=U_{23}$, and introduce the anisotropy parameter $R\equiv U'/U (<1)$. 
Using the calculations described above, we obtain $T_{\rm CSF}$ for $R=0.5$, $0.2$, and $0$ as shown in Fig.\,\ref{fig_phase_R}.  We find that $T_{\rm CSF}$ increases with decreasing $R$, although the total strength of the attractive interaction decreases for a given $|U|/t$. When the anisotropy is introduced, the trion formation is suppressed. This tendency extends the CSF phase, leading to an increase in the CSF transition temperature. 
In the limit $R\to0$, the system is equivalent to the attractive two-component Hubbard model. Superfluidity has already been observed experimentally for systems adequately described by the attractive two-component Hubbard model \cite{Chin2006}. 
Comparing the maximum CSF transition temperatures in $0  \leq R \leq 1$, 
we find that $T_{\rm CSF}^{\rm max}/t$ of the attractive three-component Hubbard model is at least roughly a half that of the two-component model. 
The findings suggest that the CSF transition temperature can be controlled via the scattering lengths, and the attractive three-component fermionic atom system in optical lattices is a candidate of observing the novel CSF and the trionic state.

In recent experiments for balanced three-component $^{6}{\rm Li}$ fermionic atoms, an anomalous increase was observed in the three-body loss at a certain magnetic field \cite{Ottenstein}. 
Its origin has been discussed in connection with the Efimov trimer states in the continuous system \cite{Naidon,Braaten,Floerchinger}.
In the low-density region of the lattice system, low-energy properties can be well described by the continuous model. 
It is thus interesting to discuss the difference and/or analogy between the Efimov trimer in the continuous system and the trionic bound state in the lattice system. 
This issue constitutes our future study.

We thank P. Naidon, Y. Takahashi, and M. Yamashita for useful comments and valuable discussions. 
Numerical computations were carried out at the Supercomputer Center, 
the Institute for Solid State Physics, University of Tokyo. 
This work was supported by a Grant-in-Aid (Grants No. 20540390 and No. 21104514) for Scientific Research from 
the Ministry of Education, Culture, Sports, Science, and Technology, Japan.



\end{document}